\documentclass[reprint, nobibnotes, amsmath, amssymb, prb, floatfix]{revtex4-2}

\usepackage{graphicx}
\usepackage{dcolumn}
\usepackage{bm}
\usepackage{xcolor}
\usepackage{mathtools}
\usepackage{lipsum}
\usepackage[colorlinks=true, linkcolor=blue, citecolor=blue, urlcolor=blue]{hyperref}
\usepackage{booktabs}
\usepackage{array}
\usepackage{multirow}
\usepackage{amsthm}

\newtheoremstyle{dashnote}% style name
  {}{}%                       space above / below (defaults)
  { }%                 body font
  {}%                         indent
  {\bfseries}%                head font
  {}%                        punctuation after head
  { }%                        space after head
  {\thmname{#1} \thmnumber{#2}\thmnote{ -  #3}}% head spec

\theoremstyle{dashnote}
\newtheorem{theorem}{Theorem}

\usepackage{physics}

\usepackage{tikz}
\usetikzlibrary{shapes.geometric, arrows.meta, positioning, calc}
\usepackage{standalone} 

\newcommand{\eps}{\varepsilon}
\newcommand{\at}[1]{\Big\vert_{#1}}
\newcommand{\wt}{\widetilde}
\let\ringaccent\r
\renewcommand{\r}{{\vb{r}}}
\DeclareRobustCommand{\AA}{\ringaccent A}

\newcommand{\G}{{\vb{G}}}

\newcommand{\rev}[1]{#1}
\begin{document}

\preprint{APS/123-QED}

\title{Rigorous foundations of the effective mass approximation: analyticity and symmetry of band extrema}
\author{Jakob Kjærulff Svaneborg}
\email{jakoks@dtu.dk}
\affiliation{CAMD, Department of Physics, Technical University of Denmark, DK-2800 Kongens Lyngby, Denmark}

\date{\today}

\begin{abstract}
The effective mass approximation is widely used across models of carrier transport, optical response, and excitons in semiconductors and insulators, but its validity hinges on the assumption that the band dispersion $E_n(\vb{k})$ at the relevant extremum is analytic. We prove that analyticity holds at any non-degenerate extremum \rev{for Kohn--Sham density functional theory (DFT) with local or hybrid exchange-correlation functionals, for Hartree--Fock, and for band-edge $G_0W_0$ quasiparticle energies in gapped systems}.
Band non-analyticity (or \emph{warping}) in these settings is therefore intrinsically tied to degeneracy.
We then use group theory to determine the symmetry-allowed form of the effective mass tensor for each of the 32 crystallographic point groups, providing a stringent consistency check on first-principles calculations. As a representative application, we show that the electron and hole effective masses at the $K$ point of monolayer MoS$_2$ must be strictly isotropic at the DFT and $G_0W_0$ levels.
\end{abstract}

\maketitle

\section{Introduction}
\label{sec:introduction}
The effective mass approximation is widely used in semiconductor physics as a simplified framework in which charge carriers in a periodic potential are described as free particles with a renormalized mass.
Often, the effective mass is introduced via $\vb{k}\cdot \vb{p}$ perturbation theory, under the implicit assumption that the band dispersion $E_n(\vb{k})$ admits a convergent Taylor expansion at the extremum of interest.
This assumption amounts to the requirement that $E_n(\vb{k})$ is real-analytic near the relevant extremum.
\rev{Analyticity of isolated bands for periodic local Schr\"odinger operators was established in the work of Kohn~\cite{kohn1959analytic} and des Cloizeaux~\cite{descloizeaux1964energy, descloizeaux1964analytical}. However, corresponding results have not, to our knowledge, been established for the nonlocal Hamiltonians of Hartree--Fock and hybrid density-functional theory (DFT), nor for $G_0W_0$ quasiparticle energies.
For these methods, the validity of the effective mass approximation in first-principles practice is generally taken for granted rather than established rigorously.
The results presented here concern static single-particle Hamiltonians with local potentials, or with nonlocal potentials whose kernels decay exponentially -- the classes for which we establish analyticity of the Bloch Hamiltonian $H(\vb{k})$ -- together with $G_0W_0$ quasiparticle energies under the hypotheses stated in Theorem~\ref{thm:analyticity}. They do not cover approaches in which the effective single-particle problem is defined self-consistently through a frequency-dependent self-energy -- such as dynamical mean-field theory (DMFT), self-consistent $GW$, or cumulant expansions -- nor dispersions renormalized by electron--phonon coupling.}
Once the validity of the effective mass approximation for a given extremum has been rigorously established, the symmetry-allowed form of the dispersion near the extremum is restricted not only by the inherent symmetry of the material, but also by the symmetry of second-order polynomials.
These constraints together determine the number of independent components of the effective mass tensor, and the directions along which they are defined; yet anisotropy is frequently discussed phenomenologically in the first-principles literature, occasionally yielding reported tensors that are inconsistent with the symmetry of the crystal. Monolayer MoS$_2$ provides a representative example: its $K$-point conduction and valence bands are non-degenerate (apart from spin) and located at a $C_{3h}$ symmetry point, which strictly forbids in-plane anisotropy of the effective mass, and yet anisotropic values have been reported~\cite{cheiwchanchamnangij2012quasiparticle, kadantsev2012electronic, xi2014tunable}.

This work addresses both the existence of a well-defined effective mass approximation and the symmetry constraints that it imposes.
In Sec.~\ref{sec:effective_mass} we state our main result on band analyticity (Theorem~\ref{thm:analyticity}) and illustrate the relationship between band warping and degeneracy with a calculation on monolayer TiO$_2$.
In Section~\ref{sec:symmetry} we tabulate the symmetry-allowed form of the effective mass tensor for each of the 32 crystallographic point groups, and apply the classification to monolayer MoS$_2$, where it also clarifies the origin of the anisotropies reported in the literature.
In section~\ref{sec:analyticity} we provide the proof of Theorem~\ref{thm:analyticity}.
Finally, in section~\ref{sec:conclusion} we summarize our results and discuss directions for future research.

\section{The effective mass approximation and band warping}
\label{sec:effective_mass}
\begin{figure*}[tbp]
  \centering
  \begin{tikzpicture}[
    font=\small,
    node distance=12mm and 6mm,
    decision/.style={
        diamond, draw=black!70, aspect=2, inner sep=1pt,
        text width=32mm, align=center, thick, minimum height=14mm,
    },
    method/.style={
        rectangle, draw=black!70, rounded corners=2pt, fill=blue!6,
        text width=36mm, align=center, thick, inner sep=4pt, minimum height=12mm,
    },
    result_ok/.style={
        rectangle, draw=black!60, rounded corners=2pt, fill=green!18,
        text width=36mm, align=center, thick, inner sep=4pt, minimum height=18mm,
    },
    result_warn/.style={
        rectangle, draw=black!60, rounded corners=2pt, fill=yellow!25,
        text width=36mm, align=center, thick, inner sep=4pt, minimum height=18mm,
    },
    result_bad/.style={
        rectangle, draw=black!60, rounded corners=2pt, fill=red!12,
        text width=34mm, align=center, thick, inner sep=4pt, minimum height=14mm,
    },
    arrow/.style={-{Stealth[length=2mm]}, thick},
    label/.style={font=\normalsize\itshape, inner sep=2pt},
]
\node[decision] (degen) {Band non-degenerate at $\vb{k}_0$?$^\ast$};
\node[result_bad, right=18mm of degen] (warped) {
    \textbf{Indeterminate}\\[2pt]
    \footnotesize Band may or may not\\
    \footnotesize be analytic
};
\node[method, below=14mm of degen, xshift=-50mm] (mlocal) {
    Local potential\\ \footnotesize (LDA, GGA, meta-GGA)
};
\node[method, below=14mm of degen] (mexch) {
    Exact exchange\\ \footnotesize (HF, hybrid DFT)
};
\node[method, below=14mm of degen, xshift=50mm] (mgw) {
    $G_0W_0$
};
\node[result_ok, below=9mm of mlocal] (rlocal) {\textbf{Analytic}};
\node[result_warn, below=9mm of mexch] (rexch) {
    \textbf{Analytic}\\[2pt] \footnotesize if gapped
};
\node[result_warn, below=9mm of mgw] (rgw) {
    \textbf{Analytic at}\\
    \textbf{band edges}\\[2pt] \footnotesize if gapped
};
\draw[arrow] (degen.south) -- ++(0,-7mm) -| (mlocal.north);
\draw[arrow] (degen.south) -- (mexch.north);
\draw[arrow] (degen.south) -- ++(0,-7mm) -| (mgw.north);
\draw[arrow] (degen.east) -- node[label, above] {no} (warped.west);
\node[label] at ($(degen.south)+(5mm,-4mm)$) {yes};
\draw[arrow] (mlocal) -- (rlocal);
\draw[arrow] (mexch) -- (rexch);
\draw[arrow] (mgw) -- (rgw);
\node[font=\small, below=5mm of rexch, text width=110mm, align=center]
    {$^\ast$Spin degeneracy in the absence of spin--orbit coupling is immaterial:\\
    Theorem~1 applies to each spin channel independently.};
\end{tikzpicture}
  \caption{Decision tree for the applicability of Theorem~1. If the band is
        analytic at $\vb{k}_0$, the effective mass tensor of Eq.~\eqref{eq:emass-tensor-definition} is
        well defined and Eq.~\eqref{eq:emass-expansion} is the leading term of a convergent
        local power series for $E_n(\vb{k})$.  The band dispersion is therefore locally described by the effective mass approximation~\eqref{eq:emass-expansion}.
        The $GW$ box's "band edges" refers to bands whose energy lies within the analyticity strip $\Omega$ defined in Sec.~\ref{subsec:analyticity}.
        For the band edge states (VBM, CBM) this is automatic in any gapped system, while for deeper bands it depends on the spectrum.
    }
  \label{fig:analyticity}
\end{figure*}
In the effective mass approximation, the band structure of a solid in the vicinity of a stationary point $\vb{k}_0$ is expressed as
\begin{equation}
\label{eq:emass-expansion}
    E_n(\vb{k}) \approx E_n(\vb{k}_0)
    + \frac{\hbar^2}{2} \sum_{ij}
    \left(\vb{k} - \vb{k}_0\right)_i M^{-1}_{ij} \left(\vb{k} - \vb{k}_0\right)_j,
\end{equation}
where the inverse effective mass tensor $M^{-1}$ is defined as
\begin{equation}
\label{eq:emass-tensor-definition}
    \left( M^{-1} \right)_{ij} \equiv \frac{1}{\hbar^2} \frac{\partial^2 E_n}{\partial k_i \partial k_j},
\end{equation}
and subscripts $(i, j)$ label the Cartesian components of the wavevector $\vb{k}$.
Equation~\eqref{eq:emass-expansion} presupposes that $E_n$ is twice differentiable at $\vb{k}_0$, so that the inverse effective 
mass tensor of Eq.~\eqref{eq:emass-tensor-definition} is well-defined, and that the right-hand side of Eq.~\eqref{eq:emass-expansion} constitutes the leading non-trivial terms of a power series expansion of $E_n$ that converges to $E_n(\vb{k})$ in a neighborhood of $\vb{k}_0$.
The latter property is precisely the defining feature of an \textit{analytic} function: $f$ is 
analytic at $\vb{k}_0$ if it admits a power series expansion that converges to 
$f(\vb{k})$ for $|\vb{k} - \vb{k}_0|$ sufficiently small.
Since an analytic function is automatically infinitely differentiable, analyticity of $E_n$ at $\vb{k}_0$ 
guarantees both requirements at once. 
The main result of the present work is the 
following theorem, which establishes such analyticity at \textit{non-degenerate} 
extrema under conditions covering the most commonly employed approximations:
\begin{theorem}[analyticity of non-degenerate bands]
\label{thm:analyticity}
Let $\vb{k}_0$ be a point in the Brillouin zone at which $E_n(\vb{k})$ is
non-degenerate. Then $E_n$ is a real-analytic function of $\vb{k}$ in a neighborhood 
of $\vb{k}_0$ in each of the following cases:
\begin{enumerate}
    \item[(i)] $H = T + V_\text{loc}(\r)$, where $V_\text{loc}$ is a local potential 
    (as in DFT with LDA, GGA, or meta-GGA exchange-correlation functionals).
    \item[(ii)] $H$ additionally contains a non-local potential whose kernel $V_\text{NL}(\vb{r},\vb{r}')$ decays exponentially with $|\vb{r}-\vb{r}'|$. This includes, in particular, exact exchange in gapped systems (Hartree--Fock or hybrid DFT).
    \item[(iii)] $E_n(\vb{k})$ is a $G_0W_0$ quasiparticle energy built on a Hamiltonian satisfying~(i) or~(ii), the system is gapped, and the energy at which the $G_0W_0$ self-energy $\Sigma$ is evaluated lies within a strip $\Omega$ around the gap, defined in section \ref{subsec:analyticity} and shown in Fig.~\ref{fig:sigma-poles}.
    % satisfies $|\eps_{n\vb{k}}^\text{QP} - \eps_{n\vb{k}}^\text{KS}| < E_\text{gap}^\text{KS}$.
\end{enumerate}
In particular, Eq.~\eqref{eq:emass-expansion} is then the leading non-trivial term 
of a convergent local power series representation of $E_n(\vb{k})$. \\
\end{theorem}
This conclusion is consistent with the numerical evidence reported in Ref.~\cite{supka2022two}, that warping appears only in the presence of degeneracies.
\rev{We note that only the degeneracy itself enters the hypothesis of Theorem~\ref{thm:analyticity}; its origin -- whether imposed by symmetry or accidental -- is immaterial.}
For such degenerate extrema, an expansion of the form~\eqref{eq:emass-expansion} often does not exist; $E_n$ then fails to be analytic at $\vb{k}_0$, and the band is said to be \emph{warped}. We use the terms \emph{warped} and \emph{non-analytic} interchangeably in the following.

For warped bands, Ref.~\cite{mecholsky2014theory} proposed the more general angular expansion
\begin{equation}
\label{eq:warped-band-dispersion}
    E_n(\vb{k}) = E_n(\vb{k}_0) + \frac{\hbar^2 k_r^2}{2 m_e} f(\theta, \phi),
\end{equation}
with $k_r = |\vb{k} - \vb{k}_0|$ and $(\theta, \phi)$ specifying the direction of $\vb{k} - \vb{k}_0$. The function $f(\theta, \phi)$ may be interpreted as a direction-dependent inverse mass, and Eq.~\eqref{eq:warped-band-dispersion} requires only one-dimensional analyticity of the band in $k_r$ at fixed $(\theta, \phi)$.
This form is therefore widely applicable, but generally harder to compute and use than the polynomial expansion~\eqref{eq:emass-expansion}, since accurate determination of $f(\theta, \phi)$ requires fine angular sampling, and reducing it to a scalar effective mass requires an application-specific angular weighting that differs between transport coefficients~\cite{mecholsky2014theory,laflamme2016precise}, density of states~\cite{mecholsky2016density,supka2022two}, and other properties.
When Eq.~\eqref{eq:emass-expansion} applies, it is therefore preferable.
This motivates the question of when, precisely, it does apply.

\begin{figure}
    \centering
    \includegraphics[width=\linewidth]{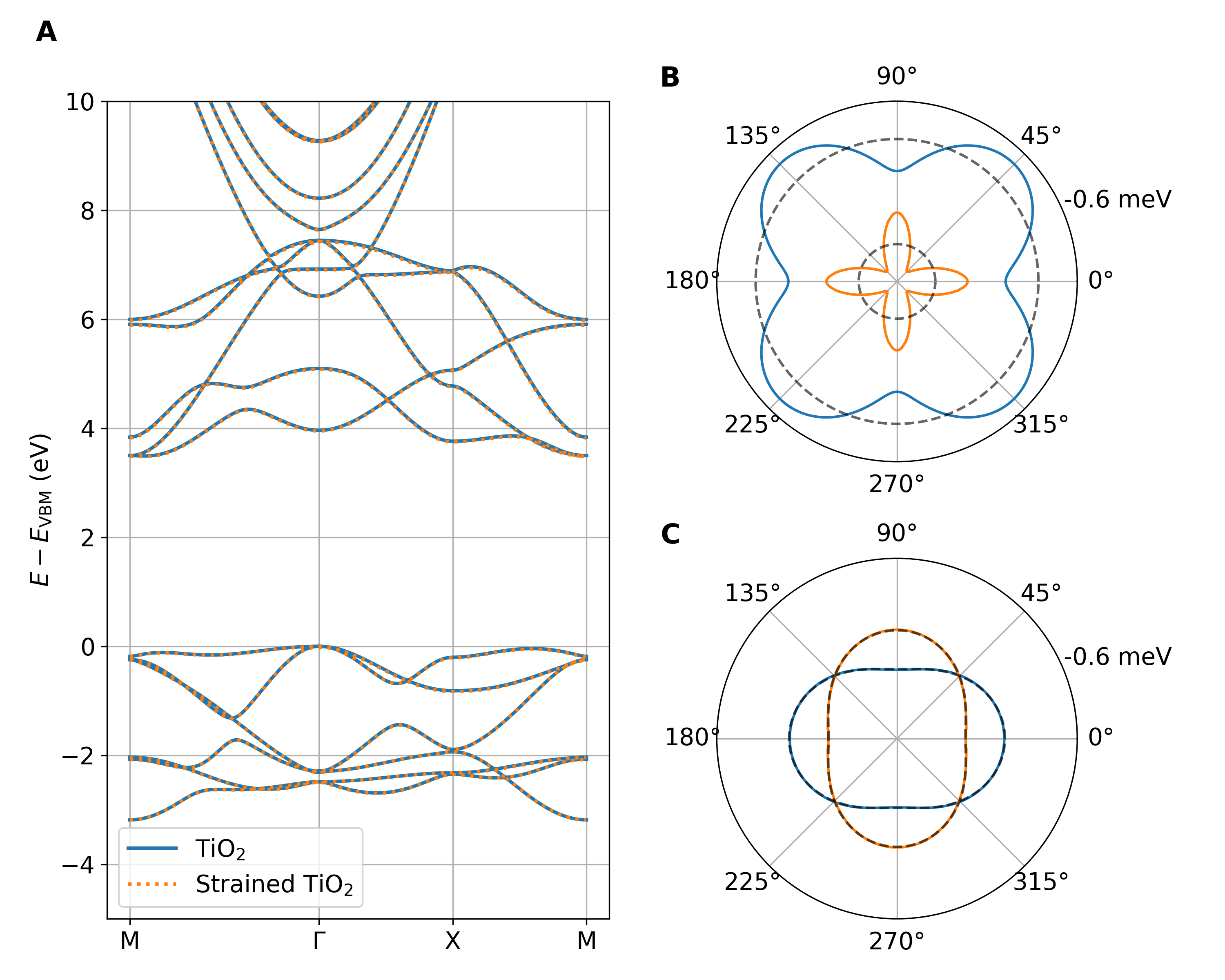}
    \caption{a) PBE band structures of relaxed and strained monolayer TiO$_2$. In the strained structure, the length of one of the in-plane crystal axes was increased by 0.1\%. This small strain reduces the symmetry of the crystal enough to lift the degeneracy of the valence bands, but is too small to have a visible effect on the band structure. 
    b, c) Inverse effective mass surfaces of the valence band maximum (VBM) of (b) relaxed  and (c) strained  monolayer TiO$_2$. In the relaxed structure, the band is degenerate and warped, and the effective mass approximation does not apply. Applying the small strain, the degeneracy is lifted, and the effective mass approximation now applies to each band separately.}
    \label{fig:tio2-warping}
\end{figure}

As an example of band warping, Fig.~\ref{fig:tio2-warping}A shows the PBE band structure of monolayer TiO${}_2$.
The valence band maximum occurs at the $\Gamma$ point and is doubly degenerate, leading to band warping in the vicinity of the degeneracy.
This is shown in Fig.~\ref{fig:tio2-warping}B where the two top valence bands are shown along a circle with radius $0.005$ \AA$^{-1}$ around the valence band maximum.
If the bands had been analytic (non-warped), the four-fold rotational symmetry of the material would dictate a completely isotropic effective mass.
The dispersion obtained by fitting the bands to Eq.~\eqref{eq:emass-expansion} is shown by the dashed circles; it is evident that they do not describe the actual band dispersion, shown in blue and orange, very well.
The degeneracy of the valence bands at $\Gamma$ may be lifted by applying a small strain of 0.1\% along the $x$-direction. 
This perturbation is so small that it has no visible effect on the band structure as shown in Fig.~\ref{fig:tio2-warping}A. However, with the degeneracy lifted, the valence bands are now analytic at $\Gamma$, and the effective mass approximation Eq.~\eqref{eq:emass-expansion} perfectly describes the bands as shown by the dashed lines in Fig.~\ref{fig:tio2-warping}C. 
We note that the strain lifts the degeneracy by reducing the 4-fold rotational symmetry to a 2-fold symmetry, thereby permitting anisotropy of the effective mass.
This anisotropy is also evident in the bands in Fig.~\ref{fig:tio2-warping}C. 
In Fig.~\ref{fig:tio2-warping}, spin-orbit coupling (SOC) is not included, and spin-up and spin-down channels are therefore degenerate. 
However, in the absence of SOC, spin degeneracy is not enough to induce non-analyticity.
Without SOC, the total electronic Hamiltonian may be written as a direct sum $H = H_\uparrow \oplus H_\downarrow$.
Because the spin-up and spin-down sectors do not couple, theorem~\ref{thm:analyticity} may be applied to the two spin channels separately.
\rev{More generally, degeneracy at an extremum is necessary for warping but not sufficient. Whenever the Hamiltonian can be written as the direct sum of two independent subfamilies, as in the spin example above, Theorem~\ref{thm:analyticity} applies to each subfamily separately, and a degeneracy between bands belonging to different subfamilies, each of which is non-degenerate within its own subfamily, does not lead to warping. In the absence of such a decomposition, we know of no general criterion for deciding whether a degenerate extremum is warped.}

\section{Symmetry Constraints}
\label{sec:symmetry}
As a physical observable, the band energy must remain invariant under all crystal symmetries. 
For three dimensions, these symmetries form a space group $\mathcal{G}$ (in two dimensions, a layer group~\cite{fu2024symmetry}). 
A symmetry operation $g = \{R_g|\bm{\tau}\} \in \mathcal{G}$ acts in real space as $\r \to R_g\r + \bm{\tau}$, and in reciprocal space as $\vb{k} \rightarrow R_g\vb{k}$, where $R_g$ is a point group operation and $\bm{\tau}$ a translation.
Invariance of the dispersion therefore requires
\begin{equation}
\label{eq:dispersion-symmetry-invariance}
   E_n(\vb{k}) = E_n(R_g^{-1}\vb{k})
\end{equation}
for any $\vb{k}$ and any $g \in \mathcal{G}$.
For a fixed $\vb{k}_0$, the subset of operations that leave $\vb{k}_0$ invariant up to a reciprocal lattice vector forms the little group $\mathcal{G}_{\vb{k}_0}$. This group is isomorphic to one of the 32 crystallographic point groups (in both two and three dimensions).
Writing $\vb{k} = \vb{k}_0 + \Delta\vb{k}$, the effective mass approximation, Eq.~\eqref{eq:emass-expansion}, depends only on $\Delta \vb{k}$.
For $h \in \mathcal{G}_{\vb{k}_0}$, Eq.~\eqref{eq:dispersion-symmetry-invariance} can be written
\begin{equation}
    E_n(\vb{k}_0 + \Delta\vb{k})
    = E_n(\vb{k}_0 + R_h^{-1}\Delta\vb{k}),
\end{equation}
showing that the dispersion in the vicinity of $\vb{k}_0$ is invariant under the action of any $h \in \mathcal{G}_{\vb{k}_0}$.
In representation-theoretic language,  the energy near $\vb{k}_0$ transforms according to the totally symmetric (trivial) representation of $\mathcal{G}_{\vb{k}_0}$.
Since the effective mass surface is a quadratic form in $\Delta \vb{k}$, we can determine the symmetry-invariant terms by decomposing this quadratic form into the irreducible representations of the relevant point group.

Such a decomposition divides the 32 crystallographic point groups into three classes, as shown in table \ref{tab:anisotropy}:
in the cubic groups, the only quadratic form transforming according to the totally symmetric representation is $(x^2 + y^2 + z^2)$. 
Consequently, for $k$-points whose little group is cubic, the effective mass tensor is necessarily isotropic.

At the opposite extreme are the lowest-symmetry groups, containing at most a 2-fold rotational axis. In these, the monomials $x^2, \ y^2$ and $z^2$ independently transform according to the trivial representation.
Any linear combination of them is therefore symmetry-allowed, and the effective-mass tensor may be fully anisotropic with three independent principal components.
The remaining groups form an intermediate class. 
Here, the trivial representation is spanned by $(x^2 + y^2)$ and $z^2$.
This dictates uniaxial symmetry, i.e. the existence of an axis such that the mass is isotropic within the plane perpendicular to the axis, but distinct along the axis.
This category includes all the non-cubic groups with an $n$-fold rotation axis with $n > 2$, and is the largest of the three classes.
These conclusions all hold both with or without SOC; although spinor states acquire a sign change under rotation by $2\pi$, the band energy $E_n(\vb{k})$ is always invariant.
The symmetry constraints on the effective mass tensor therefore depend only on the ordinary point group $\mathcal{G}_{\vb{k}_0}$.
\begin{table}
\begin{ruledtabular}
\begin{tabular}{ll}
\textbf{Fully Isotropic} 
  & $T, T_h, T_d, O, O_h$ \\

\midrule
\textbf{Uniaxially Isotropic} 
  &  $C_3, S_6, D_3, C_{3v}, D_{3d}$ \\
  &  $C_4, S_4, D_4, C_{4v}, C_{4h}, D_{2d}, D_{4h}$ \\
  &  $C_6, C_{3h}, D_6, C_{6v}, C_{6h}, D_{3h}, D_{6h}$ \\
\midrule
\textbf{Anisotropic} 
  & $C_1, C_i, C_2, C_s, C_{2h}, C_{2v}, D_2, D_{2h}$ \\
\end{tabular}
\end{ruledtabular}
\caption{Classification of the 3D crystallographic point groups by effective mass anisotropy.
These encompass also the symmetries of 2D materials.
For fully isotropic groups, the effective mass must be the same in all directions, i.e. the effective mass tensor is a multiple of the identity. 
For groups with uniaxial isotropy, there exists a high-symmetry axis such that the effective mass must be isotropic in the plane perpendicular to this axis.
Anisotropic groups permit full anisotropy in three perpendicular directions.
The table applies to systems with and without spin-orbit coupling.}
\label{tab:anisotropy}
\end{table}

These symmetry constraints provide a rigorous consistency check for first-principles calculations. As an example of their consequence, we study the electron- and hole effective masses of monolayer MoS${}_2$ calculated using the finite-difference method at the PBE level.
Monolayer MoS${}_2$ is a direct gap semiconductor with valence band maximum (VBM) and conduction band minimum (CBM) both located at the $K$-point of the Brillouin zone.
These bands are both non-degenerate at the extrema (apart from a trivial spin-degeneracy), and therefore the bands must be analytic at the PBE level according to theorem~\ref{thm:analyticity}. 
The effective mass approximation~\eqref{eq:emass-expansion} therefore applies.

\begin{figure}
    \centering
    \includegraphics[width=\linewidth]{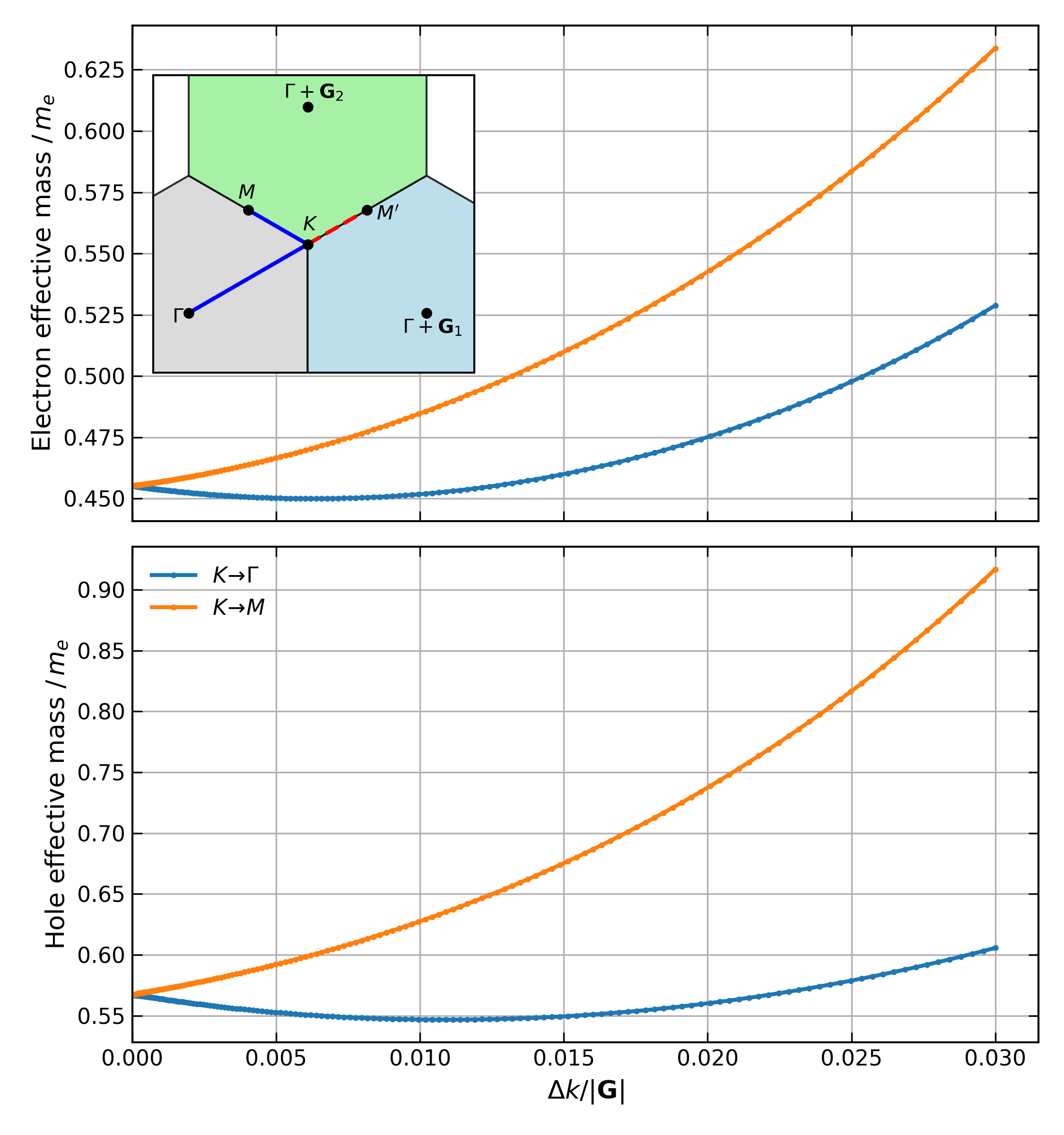}
    \caption{Convergence of the electron (top) and hole (bottom) effective masses with step size in the finite difference scheme for monolayer MoS$_2$. If too large a step size is used, an incorrect anisotropy in the effective mass is found. 
    The inset illustrates the geometry of the BZ of MoS$_{2}$ around the K-point where the VBM and CBM are found. As the M and M$'$ points are physically equivalent, the $\Gamma-$K and K$-$M directions are related by a 180$^\circ$ rotation; this provides further physical insight into the fact that the effective mass in these two directions must be equivalent for a non-warped band extremum.}
    \label{fig:mos2-emass-convergence}
\end{figure}

The little group of $K$ is isomorphic to $C_{3h}$, which according to table \ref{tab:anisotropy} signifies a uniaxially isotropic effective mass. 
There are therefore two independent directions for the effective mass - in-plane and out-of-plane components; and since we are dealing with a two-dimensional material, there is no concept of an out-of-plane effective mass. Therefore, the effective mass is fully isotropic. 
This conclusion stands in contrast to several reported calculations for monolayer MoS$_2$, which find different effective masses along the $K{-}\Gamma$ and $K{-}M$ directions or describe them as \emph{almost} isotropic~\cite{cheiwchanchamnangij2012quasiparticle, kadantsev2012electronic, xi2014tunable}.
As we now show, this apparent anisotropy is likely a numerical artifact of the finite-difference scheme.
Figure~\ref{fig:mos2-emass-convergence} shows the electron and hole effective masses of monolayer MoS${}_2$ calculated in the \rev{forward (one-sided)} finite-difference scheme without SOC \rev{(all computational details are given in Appendix~\ref{app:computational_details})}. The effective mass has been calculated both in the K$-\Gamma$ and K$-$M directions. 
As the step size is reduced, the effective mass converges to the same value regardless of the direction the limit is taken, showing that the effective mass is indeed isotropic. 
For larger step sizes, cubic terms in the dispersion introduce spurious anisotropy. 
Due to the geometry of the Brillouin zone, shown in the inset of Fig.~\ref{fig:mos2-emass-convergence}, the $K-M$ direction is equivalent to $K-M'$, which is exactly the opposite direction as $K-\Gamma$. This means that cubic terms in the dispersion contribute with opposite signs in the $K-\Gamma$ and $K-M$ directions, making this class of material particularly sensitive to step size convergence in the determination of the effective mass.
\rev{The importance of step-size convergence in effective-mass calculations has been discussed previously~\cite{laflamme2016precise}.
This issue is avoided by perturbative implementations that evaluate the band Hessian directly in the $\Delta k \to 0$ limit, such as the density-functional perturbation theory (DFPT) method of Ref.~\cite{laflamme2016precise}. Such methods may therefore be preferable when available.
The symmetry constraints derived here remain useful as an implementation-independent consistency check.}

\section{Proof of Theorem 1.}
\label{sec:analyticity}

In a periodic system, the electronic eigenstates take the form of Bloch states $\psi_{n\vb{k}}(\r) = u_{n\vb{k}}(\r) e^{i\vb{k}\cdot \r}$, where the functions $u_{n\vb{k}}$ have the periodicity of the lattice, i.e. for any lattice vector $\vb{R}$, $u_{n\vb{k}}(\r +  \vb{R}) = u_{n\vb{k}}(\r)$. 
This allows us to define the Bloch Hamiltonian $H(\vb{k})$ from the relation
\begin{equation}
\label{eq:H(k)-implicit-definition}
    \mel{\psi_{n\vb{k}}}{H}{\psi_{m\vb{k}}} \equiv \mel{u_{n\vb{k}}}{H(\vb{k})}{u_{m\vb{k}}},
\end{equation}
where $\psi_{n\vb{k}}$ may be \textit{any} Bloch-like function in the domain of $H$, i.e. it is not necessarily an eigenstate of $H$.
The real-space kernel of the Hamiltonian in Eq.~\eqref{eq:H(k)-implicit-definition} can be expressed as 
\begin{equation}
\label{eq:H(k)-real-space}
    H(\vb{k}, \vb{r}, \vb{r}') = \sum_{\vb{R}} e^{i\vb{k}\cdot\vb{R}} e^{-i\vb{k}\cdot \r} H(\vb{r}, \vb{r'}+\vb{R}) e^{i\vb{k}\cdot \r'} .
\end{equation}
If $H$ does not contain any non-local potentials, only the $\vb{R} = \vb{0}$ term contributes, and we find the well-known result       
\begin{equation}
\label{eq:local-H(k)}
    H_\text{loc}(\vb{k}, \vb{r})
= \frac{(\mathbf{p} + \hbar\mathbf{k})^2}{2m} + V(\vb{r}).\end{equation}
This is explicitly a polynomial in $\vb{k}$ and therefore analytic; in particular, as the domain of $H(\vb{k})$ is independent of $\vb{k}$, Eq.~\eqref{eq:local-H(k)} defines an analytic family of operators of type (A)~\cite{kato2013perturbation}, the most restrictive class. 
A direct consequence is that non-degenerate eigenvalues and eigenfunctions of $H_\text{loc}$ are analytic in $\vb{k}$, and we therefore have nothing more to prove in this important case, which among other models covers DFT Hamiltonians with local potentials, i.e. LDA, GGA and meta-GGA approximations\rev{~\cite{kohn1959analytic,descloizeaux1964analytical}}.

For Hamiltonians with non-local interactions, analyticity depends on the convergence of the sum \eqref{eq:H(k)-real-space}.
In particular, the Weierstrass theorem~\cite{ahlfors1979complex} guarantees that if the sum converges uniformly then $H(\vb{k}, \vb{r}, \vb{r}')$ is analytic in $\vb{k}$.
Since the non-local terms constitute a bounded perturbation of $H_\text{loc}(\vb{k})$, the full Hamiltonian remains an analytic family of type~(A).
We remind the reader that a series $f(\vb{k}) = \sum_n f_n(\vb{k})$ is said to be uniformly convergent if for all $\epsilon >0 $ there exists an $N$ such that for all $\vb{k}$ and all $N' > N$, $$\left|f(\vb{k}) - \sum_n^{N'} f_n(\vb{k})\right| < \epsilon.$$
An important example of this type is the case of exact exchange, where the Hamiltonian contains a non-local term $V_x(\r,\r') = \rho(\r, \r') V(\r,\r')$.
Here $V(\r, \r') = \frac{1}{|\r - \r'|}$ is the Coulomb interaction and 
\begin{equation}
    \rho(\r, \r') = \sum_{i \in \text{occ}} \psi_i^*(\r) \psi_i(\r')
\end{equation}
is the single-particle density matrix of the system.
For gapped materials (i.e. insulators and semiconductors) the density matrix $\rho(\r, \r')$ decays exponentially with $|\vb{r} - \vb{r'}|$ for large $|\r - \r'|$. 
This may be seen as a consequence of the principle of \textit{nearsightedness} of electronic matter~\cite{kohn1996density, prodan2005nearsightedness}. 
To be precise, exponential decay means that there exists positive constants $C, R_0$ and $\alpha$ such that
\begin{equation}
\label{eq:density-matrix-exponential-bound}
    |\rho(\r, \r')| < C e^{-\alpha |\r - \r'|}
\end{equation}
for all $(\r, \r')$ satisfying $|\r - \r'| > R_0$. 
It can be straightforwardly verified that such exponential decay is sufficient to guarantee uniform convergence of Eq.~\eqref{eq:H(k)-real-space}.
While the exchange interaction is the most commonly encountered non-local interaction, we note that the above argument holds for any non-local potential which satisfies a bound of the type Eq.~\eqref{eq:density-matrix-exponential-bound}.
\subsection{Quasiparticle energies in $G_0W_0$}
In many-body perturbation theory, the quasiparticle (QP) energies are roots of the nonlinear equation
\begin{equation}
\label{eq:GW-QP-equation}
    \omega - \eps^\text{KS}_{n\vb{k}} 
    - \Re\left[\mel{\psi_{n\vb{k}}}{\Sigma(\omega) - V^\text{xc}}{\psi_{n\vb{k}}}\right] = 0,
\end{equation}
where $\Sigma$ is the many-body self-energy. 
In standard practice, Eq.~\eqref{eq:GW-QP-equation} is \textit{linearized} around the Kohn--Sham eigenvalue $\eps^\text{KS}_{n\vb{k}}$~\cite{hybertsen1986electron, shishkin2007self, aryasetiawan1998gw}, giving the explicit formula
\begin{equation}
\label{eq:GW-QP-equation-linearized}
    \eps^\text{QP}_{n\vb{k}} = \eps^\text{KS}_{n\vb{k}}
    + Z_{n\vb{k}}\,
    \Re\mel{\psi_{n\vb{k}}}{\Sigma(\eps^\text{KS}_{n\vb{k}}) - V^\text{xc}}{\psi_{n\vb{k}}},
\end{equation}
with the QP renormalization factor
\begin{equation}
\label{eq:Z-factor}
    Z_{n\vb{k}} = \left(1 - 
    \pdv{\Re \Sigma_{n\vb{k}}}{\omega}\at{\eps^\text{KS}_{n\vb{k}}}\right)^{-1}.
\end{equation}
When $\eps^\text{QP}_{n\vb{k}}$ is found from the linearized Eq.~\eqref{eq:GW-QP-equation-linearized}, it is analytic wherever $\Sigma_{n\vb{k}}(\eps^\text{KS}_{n\vb{k}})$ and $\partial\Sigma_{n\vb{k}}/\partial\omega|_{\eps^\text{KS}_{n\vb{k}}}$ are. 
%We note that if $\Sigma_{n\vb{k}}(\omega)$ is analytic in $\omega$ at $\omega = \eps^\text{KS}_{n\vb{k}}$, this automatically implies analyticity of all its $\omega$-derivatives at that point.  
In the previous section, we established sufficient conditions for the single-particle energy $\eps^{\text{KS}}_{n\vb{k}}$ to be analytic in $\vb{k}$. 
It remains to be shown that $\Sigma_{n\vb{k}}(\omega)$ is analytic in $\vb{k}$ at $\omega = \eps^\text{KS}_{n\vb{k}}$. This is established in Sec.~\ref{subsec:analyticity}.
We note that Eq.~\eqref{eq:GW-QP-equation-linearized} is only meaningful if $Z_{n\vb{k}}$ is finite, i.e. the QP picture is valid.
We assume this in the following, and note that analyticity of $\Sigma_{n\vb{k}}(\omega)$ in $\omega$ automatically implies analyticity of $Z_{n\vb{k}}$ wherever the latter is finite.

It is also possible to solve the nonlinear QP equation~\eqref{eq:GW-QP-equation} directly rather than by linearization, which may sometimes result in better QP energies~\cite{rasmussen2021towards}.
In this case, we can establish analyticity via the implicit function theorem (IFT)~\cite{kato2013perturbation}. 
Writing
\begin{equation}
    \mathcal{F}(\vb{k}, \omega) \equiv \omega - \eps^\text{KS}_{n\vb{k}} 
    - \Re\mel{\psi_{n\vb{k}}}{\Sigma(\omega) - V^\text{xc}}{\psi_{n\vb{k}}},
\end{equation}
the QP equation~\eqref{eq:GW-QP-equation} amounts to $\mathcal{F}(\vb{k}, \omega) = 0$.
The IFT guarantees that the locus of this equation implicitly defines an analytic function $\vb{k} \to E^\text{QP}_{n\vb{k}}$ under the conditions that $\mathcal{F}$ is analytic in $(\vb{k}, \omega)$ at $\omega = E^\text{QP}_{n\vb{k}}$ and that
\begin{equation}
    \pdv{\mathcal{F}}{\omega} = 1 - \pdv{\Re \Sigma_{n\vb{k}}}{\omega} = Z_{n\vb{k}}^{-1} \neq 0.
\end{equation}
As in the linearized case, analyticity of $\mathcal{F}$ reduces to analyticity of $\Sigma_{n\vb{k}}(\omega)$ in $(\vb{k}, \omega)$ at the relevant energy; here, however, that energy is the QP energy itself rather than $\eps^\text{KS}_{n\vb{k}}$.
As we will show in the next section, this requirement is fulfilled if
\begin{equation}
\label{eq:qp-correction-must-be-smaller-than-gap}
    \left|\eps^\text{QP}_{n\vb{k}} - \eps^\text{KS}_{n\vb{k}}\right| < E^\text{KS}_\text{gap},
\end{equation}
which is typically satisfied in practice.

\subsection{Analyticity of the $G_0W_0$ self-energy}
\label{subsec:analyticity}
The time-ordered $G_0W_0$ self-energy which enters in the quasi-particle equation is
\begin{equation}
\label{eq:sigma-rr}
  \Sigma(\mathbf{r},\mathbf{r}';\omega)
  = \frac{i}{2\pi}\int_{-\infty}^{\infty} d\omega'\;
    G_0(\mathbf{r},\mathbf{r}';\omega{+}\omega')\, W(\mathbf{r},\mathbf{r}';\omega'),
\end{equation}
where
\begin{equation}
\label{eq:G-spectral-representation}
G_0(\mathbf{r},\mathbf{r}';\omega) 
= \sum_n \frac{\phi_n(\r) \phi_n^*(\r')}
{\omega - \eps_n + i \eta \ \text{sgn}(\eps_n - \mu)}
\end{equation}
is the single-particle Green's function, and $W = \epsilon^{-1}V$ the screened Coulomb interaction.
Introducing $\Sigma(\vb{k})$ in the same manner as $H(\vb{k})$ of Eq.~\eqref{eq:H(k)-implicit-definition}, we obtain a Fourier-like series

\begin{equation}
\label{eq:sigma-lattice-fourier}
  \Sigma_{n\mathbf{k}}(\omega)
  = \sum_{\mathbf{R}} e^{i\mathbf{k}\cdot\mathbf{R}}\;
    \sigma_{n\vb{k}}(\vb{R},\omega),
\end{equation}
where
\begin{widetext}
\begin{equation}
\label{eq:sigma-R}
  \sigma_{n\vb{k}}(\vb{R},\omega)
  = \int_{\mathrm{cell}} d\boldsymbol{\tau}
    \int_{\mathrm{cell}} d\boldsymbol{\tau}'\;
    u_{n\mathbf{k}}^*(\boldsymbol{\tau})\;
    e^{i\mathbf{k}\cdot(\boldsymbol{\tau}'-\boldsymbol{\tau})}\;
    \Sigma(\boldsymbol{\tau},\,\mathbf{R}{+}\boldsymbol{\tau}';\omega)\;
    u_{n\mathbf{k}}(\boldsymbol{\tau}').
\end{equation}
\end{widetext}
If the single-particle Hamiltonian is analytic in $\vb{k}$, as discussed in the previous section, and $\eps_{n\vb{k}}$ is non-degenerate, then the wavefunction $u_{n\vb{k}}$ is locally analytic in $\vb{k}$~\cite{kato2013perturbation}.
Each $\sigma_{n\vb{k}}(\vb{R}, \omega)$ is analytic in $\vb{k}$ as it is defined from a finite-volume integral of functions that are analytic in $\vb{k}$.
We now examine the conditions under which $\sigma_{n\vb{k}}(\vb{R},\omega)$ decays exponentially with $|\vb{R}|$; as noted in the previous section, this condition is sufficient for the sum~\eqref{eq:sigma-lattice-fourier} to define an analytic function by the Weierstrass theorem.

\begin{figure}
    \centering
    \includegraphics[width=\linewidth]{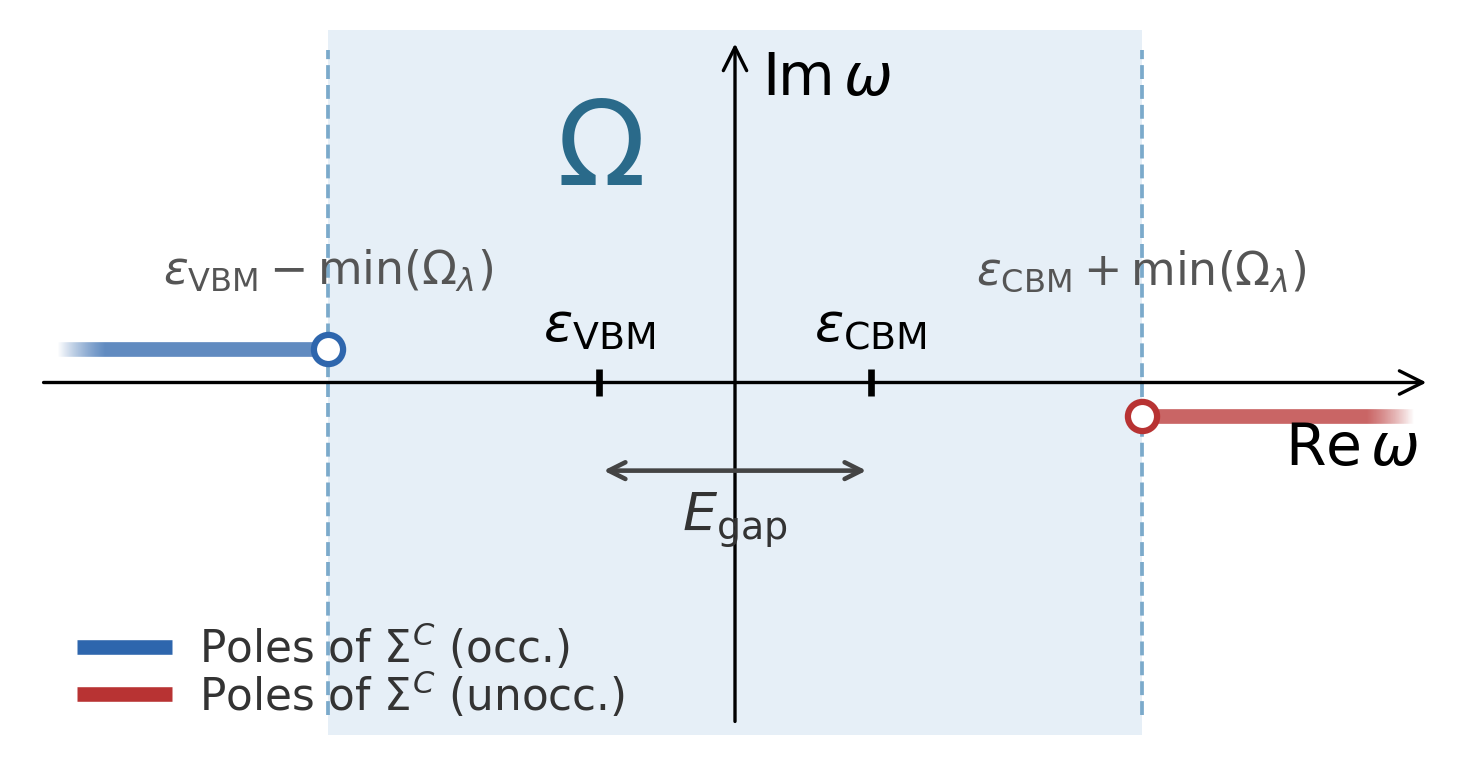}
    \caption{Spectral structure of the correlation self-energy $\Sigma^C$ in the $GW$ approximation. The self-energy is analytic in $\omega$ in closed subsets of the region $\Omega$ in which it has no poles.}
    \label{fig:sigma-poles}
\end{figure}

We focus on the correlation part $\Sigma^C = G \overline{W}$ of the $GW$ self-energy, where $\overline{W} = W - V$, because  the exchange part $\Sigma^X = GV$ is already known to be exponentially localized due to the previously discussed localization of the density matrix in gapped systems.
The correlation self-energy $\Sigma^C$ may be obtained from Eq.~\eqref{eq:sigma-rr} by simply replacing $W$ by $\overline{W}$.
To proceed, we seek a formal spectral representation of $\Sigma^C$. 
This can be achieved by noting that $\overline{W}$ admits the general spectral representation~\cite{holzer2019gw}
\begin{equation}
\label{eq:W-spectral-representation}
    \overline{W}(\r, \r' ; \omega) = \sum_\lambda
    \frac{w^*_\lambda(\r)  w_\lambda(\r')}{\omega - \Omega_\lambda + i \eta}
    -
    \frac{w_\lambda(\r)  w^*_\lambda(\r')}{\omega + \Omega_\lambda - i \eta},
\end{equation}
where the poles $\Omega_\lambda$ correspond to neutral excitation energies.
Combining Eqs.~\eqref{eq:sigma-rr},~\eqref{eq:G-spectral-representation} and \eqref{eq:W-spectral-representation}, we arrive at the spectral representation of $\Sigma^C$~\cite{duchemin2020robust}:
\begin{equation}
\label{eq:SigmaC-spectral-rep}
\begin{aligned}
\Sigma^C(\r, \r'; \omega) = 
\sum_\lambda \Bigg[ \sum_{i \in \text{occ}} 
\frac{w^*_\lambda(\r) w_\lambda(\r') \phi_i(\vb{r}) \phi^*_i(\vb{r}')}
{\omega - \eps_i + \Omega_\lambda - i \eta} \\
+  \sum_{a \in \text{unocc}} 
\frac{w_\lambda(\r) w^*_\lambda(\r') \phi_a(\vb{r}) \phi^*_a(\vb{r}')}
{\omega - \eps_a - \Omega_\lambda + i \eta} \Bigg].
\end{aligned}
\end{equation}
We note that $\Sigma^C$ has poles at $\eps^\text{KS}_i - \Omega_\lambda + i\eta$ for occupied states $i$ and at $\eps^\text{KS}_a + \Omega_\lambda - i\eta$ for unoccupied states $a$. Since the $\Omega_\lambda$ are strictly positive, this defines a strip
\begin{equation}
\label{eq:strip-definition}
    \Omega = \left\{\omega : \Re(\omega) \in 
    \left(E^\text{KS}_\text{VBM} - \min_\lambda \Omega_\lambda,\;
    E^\text{KS}_\text{CBM} + \min_\lambda \Omega_\lambda\right)\right\}
\end{equation}
around the band gap in which $\Sigma^C(\omega)$ is analytic, as illustrated in Fig.~\ref{fig:sigma-poles}. If $W$ is calculated in the RPA, the $\Omega_\lambda$ are bounded below by the KS gap, so the strip is at least as wide as $E^\text{KS}_\text{VBM} - E^\text{KS}_\text{gap} < \Re(\omega) < E^\text{KS}_\text{CBM} + E^\text{KS}_\text{gap}$.
In Appendix \ref{app:representation_of_sigma_c_from_resolvents_of_two_auxiliary_operators} we show how $\Sigma^C$ can be represented in terms of the resolvents of two families of auxiliary operators $M^{i/a}_\lambda$. 
Using this representation we demonstrate via the Combes--Thomas estimate~\cite{combes1973asymptotic} that $\Sigma^C$ is exponentially localized in $|\vb{r} - \vb{r'}|$ for $\omega \in \Omega$. 
As in the case of exact exchange, this localization guarantees uniform convergence of the series~\eqref{eq:sigma-lattice-fourier}, and by the Weierstrass theorem, analyticity of $\Sigma_{n\vb{k}}(\omega)$ in both $\vb{k}$ and $\omega$ for $\omega \in \Omega$ whenever $\psi_{n\vb{k}}$ is non-degenerate.
Since $\eps^\text{KS}_{n\vb{k}}$ lies in $\Omega$ for any band-edge state of a gapped system, the explicit linearized formula~\eqref{eq:GW-QP-equation-linearized} establishes analyticity of $\eps^\text{QP}_{n\vb{k}}$ in $\vb{k}$ directly.
If the nonlinear QP equation~\eqref{eq:GW-QP-equation} is used, analyticity holds if $E^\text{QP}_{n\vb{k}} \in \Omega$. 
When $W$ is calculated in the RPA, as typically done, this guarantees analycity as long as the QP correction is smaller than the gap, i.e. Eq.~\eqref{eq:qp-correction-must-be-smaller-than-gap}.
The effective mass approximation therefore holds at non-degenerate band edges of gapped semiconductors and insulators at the $G_0W_0$ level as long as the $G_0W_0$ calculation is based on a single-particle Hamiltonian satisyfing either cases \textit{(i)} or \textit{(ii)} of Theorem~\ref{thm:analyticity}.
We note that the analyticity applies to all non-degenerate bands whose energy lies within the strip $\Omega$. 
This distinguishes $G_0W_0$ from case~(ii) of Theorem~\ref{thm:analyticity}, where all non-degenerate bands are analytic.

\section{Conclusion}
\label{sec:conclusion}
We have established conditions under which the effective mass approximation, Eq.~\eqref{eq:emass-expansion}, is rigorously valid at non-degenerate band extrema in semiconductors and insulators.
Specifically, $E_n(\vb{k})$ is analytic at any point $\vb{k}_0$ where it is non-degenerate for the standard mean-field Hamiltonians (DFT with local exchange-correlation potentials, Hartree--Fock, and hybrid DFT).
The same applies to non-degenerate band energies of gapped systems at the $G_0W_0$ level when $\Sigma$ is evaluated within the strip $\Omega$ surrounding the gap. 
If the linearized QP equation~\eqref{eq:GW-QP-equation-linearized} is used, this condition is always satisfied at the band edges.
In contrast, if the nonlinear QP equation~\eqref{eq:GW-QP-equation} is used, an additional condition $|\eps^\text{QP}_{n\vb{k}} - \eps^\text{KS}_{n\vb{k}}| < E^\text{KS}_\text{gap}$ is necessary.

This implies that band warping at a non-degenerate extremum is impossible in these settings, so warping is intrinsically tied to degeneracy.

For analytic extrema, the symmetry-allowed form of the effective mass tensor is fully determined by the little group $\mathcal{G}_{\vb{k}_0}$ of $\vb{k}_0$, and the decomposition of quadratic forms into its irreducible representations yields three classes of materials which permit different degrees of effective mass anisotropy. These findings are summarized in Table~\ref{tab:anisotropy}, and provide a stringent consistency check for first-principles calculations.
\rev{This also suggests a practical application: effective masses reported in high-throughput databases~\cite{ricci2017ab, haastrup2018computational, gjerding2021recent} can be checked for consistency with the crystal symmetry at essentially no computational cost, flagging entries whose mass tensors are incompatible with the little group of the extremum. We leave a systematic application of this check to future work.}

Several extensions of the present results would be of interest.
The proof does not extend straightforwardly to the self-consistent variants of $GW$. 
In eigenvalue self-consistent $GW$~\cite{shishkin2007self} (ev$GW$) and QP self-consistent $GW$~\cite{van2006quasiparticle} (qs$GW$), self-consistency is obtained via the construction of a modified single-particle Hamiltonian (explicitly in qs$GW$; implicitly in ev$GW$). 
These auxiliary Hamiltonians may not be exponentially localized, even if the initial $H_0$ is, and our proof of analyticity of $H(\vb{k})$ therefore does not apply. 
In fully self-consistent $GW$ (sc$GW$), by contrast, the Green's function is no longer the resolvent of any single-particle Hamiltonian, and the spectral representation of $\Sigma^C$ underlying our localization argument is therefore invalidated.
\rev{Similar obstructions apply to other approaches in which the effective single-particle problem involves a frequency-dependent self-energy, such as DMFT and cumulant-expansion spectral functions, and to band structures renormalized by electron--phonon coupling, including zero-point renormalization; these cases lie outside the scope of the present results.}

Finally, degenerate extrema are known to be sometimes -- but not always -- warped, and the conditions distinguishing the two cases would benefit from a similarly systematic treatment.

\section*{Acknowledgments}
The author thanks Kristian Sommer Thygesen for many useful discussions, and Jan Philip Solovej for helpful comments on an earlier version of the manuscript.

\appendix

\section{\rev{Computational details}}
\label{app:computational_details}

\rev{All density-functional theory calculations were performed with the GPAW electronic-structure code~\cite{mortensen2024gpaw} in plane-wave mode, using the PBE exchange--correlation functional~\cite{perdew1996generalized} and the Atomic Simulation Environment (ASE)~\cite{larsen2017atomic}. The relaxed structures of monolayer TiO$_2$ and MoS$_2$ were taken from the C2DB database~\cite{haastrup2018computational, gjerding2021recent}. Plane-wave cutoffs of 800~eV (TiO$_2$) and 1400~eV (MoS$_2$) were used, together with $\Gamma$-centered $k$-point grids of density 12 points per \AA$^{-1}$, Fermi--Dirac occupation smearing of 0.01~eV, and 15--18~\AA{} of vacuum separating the periodic images of the layers. Spin--orbit coupling was not included. The band energies entering Figs.~\ref{fig:tio2-warping} and \ref{fig:mos2-emass-convergence} were computed non-self-consistently from the converged ground-state density, with the eigenstates converged to $10^{-10}$~eV$^2$ per electron.}

\rev{For the strained TiO$_2$ calculation of Fig.~\ref{fig:tio2-warping}, one in-plane lattice vector was lengthened by 0.1\% with the internal coordinates held fixed at their values in the unstrained, relaxed structure (clamped ions); the purpose of this calculation is to demonstrate the lifting of the band degeneracy by the symmetry reduction of the lattice. The inverse-mass surfaces of Fig.~\ref{fig:tio2-warping} show the direction-dependent inverse mass [the function $f$ of Eq.~\eqref{eq:warped-band-dispersion}], obtained by sampling the band energies on a circle of radius $k_r = 0.005\,|\G|$ around the extremum -- where $|\G|$ is the common length of the two orthogonal in-plane reciprocal lattice vectors -- in 500 equally spaced directions $\hat{\vb{u}}$, and estimating the inverse mass along each direction from the two-point difference $2\left[E(\vb{k}_0 + k_r \hat{\vb{u}}) - E(\vb{k}_0)\right]/(\hbar^2 k_r^2)$.}

\rev{The effective masses of Fig.~\ref{fig:mos2-emass-convergence} were obtained from the one-sided (forward) three-point finite-difference estimator
\begin{equation}
\label{eq:forward-difference-estimator}
    \frac{1}{m(\Delta k, \hat{\vb{u}})} = \frac{E(\vb{k}_0 + 2\Delta k\, \hat{\vb{u}}) - 2E(\vb{k}_0 + \Delta k\, \hat{\vb{u}}) + E(\vb{k}_0)}{\hbar^2\, \Delta k^2},
\end{equation}
evaluated independently at each of 141 step sizes $\Delta k$ between $10^{-4}\,|\G|$ and $0.03\,|\G|$, with $\hat{\vb{u}}$ the unit vector along the $K$--$\Gamma$ or $K$--$M$ direction. Note that the estimator samples the band up to a distance $2\Delta k$ from the extremum.}

\rev{The input structures, raw band energies, and analysis scripts underlying Figs.~\ref{fig:tio2-warping} and \ref{fig:mos2-emass-convergence} are openly available on Zenodo~\cite{svaneborg2026supporting}.}

\section{Representation of $\Sigma^C$ from resolvents of two auxiliary operators}
\label{app:representation_of_sigma_c_from_resolvents_of_two_auxiliary_operators}
Using the projector into the occupied subspace $P_\text{occ} = \sum_{i \in \text{occ}} \ketbra{\psi_i}$, whose real-space representation is the density matrix $\rho(\r, \r')$, we now introduce two families of operators $M^{i/a}_\lambda$ given by
\begin{equation}
\label{eq:Mi}
    M^i_\lambda = \left(H_\text{KS} - \Omega_\lambda\right)P_\text{occ} + C(1 - P_\text{occ})
\end{equation}
and
\begin{equation}
\label{eq:Ma}
    M^a_\lambda = \left(H_\text{KS} + \Omega_\lambda\right) \left(1 - P_\text{occ}\right) + C' P_\text{occ},
\end{equation}
where $C$ and $C'$ are arbitrary constants.
We wish to show that $\Sigma^C$ can be written in terms of the resolvents
$R^{i/a}_\lambda(\omega) = (\omega - M^{i/a}_\lambda)^{-1}$. 
These have the form
\begin{equation}
\label{eq:resolvent-i}
R^i_\lambda(\omega, \vb{r}, \vb{r}') = \frac{\delta(\vb{r}-\vb{r'}) - \rho(\vb{r}, \vb{r'})}{\omega - C} + \sum_{i \in \text{occ}} \frac{\phi_i(\vb{r}) \phi^*_i(\vb{r}')}
{\omega - \eps_i + \Omega_\lambda},
\end{equation}
and 
\begin{equation}
\label{eq:resolvent-a}
R^a_\lambda(\omega, \vb{r}, \vb{r}') = \frac{\rho(\vb{r}, \vb{r}')}{\omega - C'} + \sum_{a \in \text{unocc}} \frac{\phi_a(\vb{r}) \phi^*_a(\vb{r}')}
{\omega - \eps_a - \Omega_\lambda},
\end{equation}
which may be seen from the identity $R^{i/a}_\lambda(\omega)(\omega - M^{i/a}_\lambda) \equiv 1$, which can be verified to hold on the occupied and unoccupied subspaces independently. 
From Eqs. \eqref{eq:resolvent-i} and \eqref{eq:resolvent-a}, the spectral representation of $\Sigma^C$ can be written
\begin{widetext}
\begin{equation}
\label{eq:SigmaC-resolvent}
\begin{aligned}
\Sigma^C(\r, \r'; \omega) = 
\sum_\lambda w^*_\lambda(\r) w_\lambda(\r') \left(R^i_\lambda(\r, \r'; \ \omega - i \eta)
- \frac{\delta(\vb{r}-\vb{r'}) - \rho(\vb{r}, \vb{r'})}{\omega - C - i \eta} \right) \\
+\sum_\lambda  w_\lambda(\r) w^*_\lambda(\r') \left(R^a_\lambda(\r, \r'; \ \omega + i \eta) - \frac{\rho(\vb{r},\vb{r'})}{\omega - C' + i \eta}\right).
\end{aligned}
\end{equation}
\end{widetext}
The shifts $C/C'$ are gauge parameters: they fix the eigenvalues of $M^{i/a}_\lambda$ on the unoccupied/occupied subspaces to $C/C'$, and the corresponding poles of $R^{i/a}_\lambda$ cancel exactly against the subtracted terms in Eq.~\eqref{eq:SigmaC-resolvent}.
Hence $\Sigma^C$ is independent of $C$ and $C'$, and has no pole inside the gap. 
We now take $C, C' \notin \Omega$, so that resolvents $R^{i/a}_\lambda$ are bounded on all of $\Omega$.

In Appendix~\ref{app:combes-thomas}, we show that the Combes--Thomas estimate~\cite{combes1973asymptotic} can be applied to the operators $M^{i/a}_\lambda$.
For $\omega \in \Omega$, this implies that the resolvents $R^{i/a}_\lambda$ are exponentially localized;
precisely, this means that the localized operator norm $\norm{\chi_\mathbf{0} R_\lambda^{i/a}(\omega)\chi_{\vb{R}}}$ decays exponentially in $|\vb{R}|$ for $|\vb{R}|$ sufficiently large, where $\chi_\mathbf{0}$ and $\chi_\mathbf{R}$ are characteristic functions of unit cells at the origin and at lattice vector $\mathbf{R}$, respectively.
Such localized matrix elements are precisely what enters in the definition of the $\sigma_{n\vb{k}}(\vb{R}, \omega)$ in Eq.~\eqref{eq:sigma-R}.
Since $w_\lambda(\r)$ and $w_\lambda(\r')$ enter Eq.~\eqref{eq:SigmaC-resolvent} as local multiplicative prefactors, the exponential decay carries over to $\sigma_{n\vb{k}}(\vb{R}, \omega)$.

\section{Combes--Thomas estimate for non-local Hamiltonians}
\label{app:combes-thomas}

We show that the operators $M^{i/a}_\lambda$ of Eqs.~(\ref{eq:Mi}--\ref{eq:Ma}) have resolvents whose matrix elements between spatially separated unit cells decay exponentially with the separation.
Let $\Omega_\mathbf{0}$ and $\Omega_\mathbf{R}$ denote unit cells centered at the origin and at a lattice vector $\vb{R}$ respectively, and let $\chi_\mathbf{0}$ and $\chi_\mathbf{R}$ be the corresponding characteristic functions, i.e.
\begin{equation}
    \chi_{\vb{R}}(\r) = \begin{cases}
    1 \qq{if} \r \in \Omega_{\vb{R}}, \\
    0 \qq {otherwise.}
    \end{cases}
\end{equation}
We wish to establish bounds of the form
\begin{equation}
\label{eq:ct-goal}
    \norm{ \chi_\mathbf{0}\, (\omega - M)^{-1}\, \chi_\mathbf{R} } \leq C\, e^{-\alpha\, |\vb{R}|}
\end{equation}
for constants $C, \alpha > 0$, $\omega$ outside the spectrum of $M$, and $|\vb{R}|$ sufficiently large.
The bound~\eqref{eq:ct-goal} is the key needed to establish analyticity of the $GW$ self-energy in Eq.~\eqref{eq:sigma-lattice-fourier} since the terms $\sigma_{n\vb{k}}(\vb{R}, \omega)$ in the sum are matrix elements of $\Sigma$ between functions localized in the unit cells $\Omega_\mathbf{0}$ and $\Omega_\mathbf{R}$.
The estimate~\eqref{eq:ct-goal} therefore implies $|\sigma_{n\vb{k}}(\vb{R}, \omega)| \leq C' e^{-\alpha |\vb{R}|}$ for large $|\vb{R}|$, which guarantees uniform convergence of the lattice sum and thus analyticity of the self-energy in $\vb{k}$.

We first prove the estimate for bounded operators with exponentially decaying kernels. 
In general, however, the Hamiltonian of an electronic system is unbounded due to the Laplacian in the kinetic energy term. 
We therefore extend the argument to Schr\"odinger-like Hamiltonians that may contain an additional bounded, non-local (but exponentially decaying) potential.

The two operators $M^{i/a}_\lambda$ require different treatments.
The operator $M^i_\lambda$ of Eq.~\eqref{eq:Mi} is bounded due to the projection into the occupied subspace. On the other hand, $M^a_\lambda$ (Eq.~\eqref{eq:Ma}) can be written
\begin{equation}
    M^a_\lambda = \left(H_\text{KS} + \Omega_\lambda\right)
    - \left(H_\text{KS} + \Omega_\lambda\right)P_\text{occ}
    + C' P_\text{occ},
\end{equation}
i.e.\ a shifted Kohn--Sham Hamiltonian plus a bounded perturbation,
and is treated by the unbounded-operator extension in the following.

\subsection*{Bounded operators}

Let $M$ be a bounded operator with an exponentially decaying kernel, i.e.\ $|M(\r, \r')| \leq C e^{-\mu |\r - \r'|}$ for $|\r - \r'|$ larger than some $R_0$.
We denote the resolvent $R_M(\omega) = (\omega - M)^{-1}$.

Following the standard Combes--Thomas approach, we define $U_\alpha = e^{\alpha \vb{a} \cdot \r}$ for a unit vector $\vb{a}$ and $\alpha > 0$, and the conjugated operator $\wt{M}_\alpha = U_\alpha M U_\alpha^{-1}$.
Using the identity $U A^{-1} U^{-1} = (U A U^{-1})^{-1}$, the resolvent of $M$ can be written
\begin{equation}
\label{eq:conjugated-resolvent}
    R_M(\omega) = U_\alpha^{-1}\, (\omega - \wt{M}_\alpha)^{-1}\, U_\alpha.
\end{equation}
Inserting characteristic functions $\chi_\mathbf{0}$ and $\chi_\mathbf{R}$ on either side and taking operator norms, we find
\begin{equation}
\label{eq:ct-factored}
    \norm{ \chi_\mathbf{0}\, R_M(\omega)\, \chi_\mathbf{R} } 
    \leq \norm{ \chi_\mathbf{0}\, U_\alpha^{-1} } \;
    \norm{ (\omega - \wt{M}_\alpha)^{-1} } \;
    \norm{ U_\alpha\, \chi_\mathbf{R} }.
\end{equation}
The operators $\chi_\mathbf{0}\, U_\alpha^{-1}$ and $U_\alpha\, \chi_\mathbf{R}$ are multiplicative operators restricted to bounded domains, and are therefore bounded.
Their norms depend on the direction $\vb{a}$. 
We choose $\vb{a} = -\hat{\vb{R}}$ with $\hat{\vb{R}} = \vb{R} / |\vb{R}|$ and define $d = \sup_{\r \in \Omega_\mathbf{0}} |\r|$. 
For $\chi_\mathbf{0}\, U_\alpha^{-1}$, the norm is then $\sup_{\r \in \Omega_\mathbf{0}}\{e^{\alpha \hat{\vb{R}} \cdot \r}\} \leq e^{\alpha d}$.
For $U_\alpha\, \chi_\mathbf{R}$, we can similarly bound $\norm{U_\alpha \chi_\mathbf{R}} \leq e^{-\alpha(|\vb{R}| - d)}$.
Using these bounds in Eq.~\eqref{eq:ct-factored}, we obtain
\begin{equation}
\label{eq:ct-bounded}
    \norm{ \chi_\mathbf{0}\, R_M(\omega)\, \chi_\mathbf{R} } 
    \leq C_\alpha\, e^{-\alpha(|\vb{R}| - 2d)},
\end{equation}
where $C_\alpha = \norm{ (\omega - \wt{M}_\alpha)^{-1} }$.
For $|\vb{R}|$ much larger than the cell size $d$, this gives the desired exponential decay in $|\vb{R}|$. 
However, the above argument works only provided the conjugated resolvent $(\omega - \wt{M}_\alpha)^{-1}$ exists and is bounded.
We now show that for sufficiently small $\alpha$, this is indeed the case.

The so-called Schur test bounds the norm of an operator $K$ by integrals of its matrix elements, namely
\begin{equation}
    \norm{K}^2 \leq \sup_{\vb{\r}}\int \dd \r' |K(\r, \r')| \cdot \sup_{\vb{\r'}}\int \dd \r |K(\r, \r')|.
\end{equation}
Consider the matrix elements of $\wt{M}_\alpha - M$,
$$(\wt{M}_\alpha - M)(\r, \r') = (e^{\alpha \vb{a}\cdot(\r - \r')} - 1)\, M(\r, \r').$$
For $|\r - \r'| > R_0$, the exponential decay of $M$ means that the matrix elements of $(\wt{M}_\alpha - M)$ decay as $e^{(\alpha - \mu)|\r - \r'|}$, which is integrable for $\alpha < \mu$. 
Meanwhile, the region where $|\r - \r'| \leq R_0$ is finite in size, and the integral is therefore finite as long as $M$ contains no non-integrable singularities, which is true of the operators $M^{i/a}_\lambda$ in the main text.
From these considerations, it can be immediately verified that the Schur test guarantees that $\norm{\wt{M}_\alpha - M} \to 0$ as $\alpha \to 0$.

The resolvent of $\wt{M}_\alpha$ can be written
$$(\omega - \wt{M}_\alpha)^{-1} = R_M(\omega)\left[1 - (\wt{M}_\alpha - M)R_M(\omega)\right]^{-1}.$$
For $\omega$ in the resolvent set of $M$, $R_M(\omega)$ is bounded.
We now choose $\alpha$ sufficiently small such that $\norm{(\wt{M}_\alpha - M) R_M(\omega)} < 1$. 
Then, we may write
\begin{equation}
    \left[1 - (\wt{M}_\alpha - M)R_M(\omega)\right]^{-1} 
    = \sum_{n=0}^\infty \left[(\wt{M}_\alpha - M)R_M(\omega)\right]^n,
\end{equation}
where the bound on the operator norm of the terms in the sum guarantees convergence of the series, and therefore establishes that the left-hand side is bounded. 
This shows that for sufficiently small $\alpha$ and $\omega$ in the resolvent set of $M$, $(\omega - \wt{M}_\alpha)^{-1}$ exists and is bounded. This completes the proof of~\eqref{eq:ct-goal}.

\subsection*{Unbounded Hamiltonians with non-local potentials}
We now extend the estimate to unbounded operators of the form
\begin{equation}
\label{eq:H-decomposition}
    H = \underbrace{-\nabla^2 + V_\text{loc}}_{H_\text{loc}} + V_\text{nl},
\end{equation}
where $\nabla^2$ is the Laplacian, $V_\text{loc}$ is a local potential, and $V_\text{nl}$ is a non-local but exponentially decaying potential, i.e.\ $|V_\text{nl}(\r, \r')| \leq C_V e^{-\mu |\r - \r'|}$ for $|\r - \r'|$ sufficiently large.
For our application, $M^a_\lambda$ falls into this framework with
$$V_\text{nl} = -\left(H_\text{KS} + \Omega_\lambda\right)P_\text{occ} + C' P_\text{occ}.$$

The structure of the proof is the same: we conjugate $H$ by $U_\alpha = e^{\alpha \vb{a}\cdot \r}$ and show that the conjugated resolvent remains bounded for small enough $\alpha$. The bound~\eqref{eq:ct-goal} then follows from the same factorization as in Eq.~\eqref{eq:ct-factored}.

The perturbation $\wt{H}_\alpha - H$ decomposes as
\begin{equation}
    \wt{H}_\alpha - H 
    = \underbrace{\left(\wt{H}_{\text{loc},\alpha} - H_\text{loc}\right)}_{W_\alpha^\text{loc}}
    + \underbrace{\left(\wt{V}_{\text{nl},\alpha} - V_\text{nl}\right)}_{W_\alpha^\text{nl}}.
\end{equation}

Consider first the local part. Since $V_\text{loc}$ commutes with $U_\alpha$, the perturbation acts only on the kinetic energy,
\begin{equation}
    U_\alpha (-\nabla^2) U_\alpha^{-1} = -(\nabla - \alpha \vb{a})^2 = -\nabla^2 + 2\alpha \vb{a} \cdot \nabla - \alpha^2,
\end{equation}
so $W_\alpha^\text{loc} = 2\alpha \vb{a} \cdot \nabla - \alpha^2$.
Since $\nabla$ is relatively bounded with respect to $\nabla^2$ (and hence with respect to $H$), for any $\psi$ in the domain $\mathcal{D}$ of $H$, we have
\begin{equation}
\label{eq:rel-bound}
    \norm{ W_\alpha^\text{loc} \psi } 
    \leq 2\alpha \left(a_0 \norm{ H \psi } + b_0 \norm{ \psi }\right) + \alpha^2 \norm{ \psi },
\end{equation}
where $a_0$ and $b_0$ are constants such that $\norm{ \nabla \psi } \leq a_0 \norm{ H \psi } + b_0 \norm{ \psi }$.
This is the standard step in the Combes--Thomas argument for Schr\"odinger operators and does not require $H$ to be bounded~\cite{combes1973asymptotic}.

The non-local part $W_\alpha^\text{nl}$ has the same structure as in the bounded case, with kernel $(e^{\alpha \vb{a} \cdot (\r - \r')} - 1) V_\text{nl}(\r, \r')$.
Since $V_\text{nl}$ is bounded with an exponentially decaying kernel, the Schur test gives 
$\norm{ W_\alpha^\text{nl} } \to 0$ as $\alpha \to 0$ for $\alpha < \mu$.

As in the previous section, we may write
\begin{equation}
    (\omega - \wt{H}_\alpha)^{-1} = R_H(\omega) \left[1 - (W_\alpha^\text{loc} + W_\alpha^\text{nl}) R_H(\omega)\right]^{-1},
\end{equation}
where $R_H(\omega) = (\omega - H)^{-1}$.
The right-hand side exists and is bounded when $\norm{(W_\alpha^\text{loc} + W_\alpha^\text{nl}) R_H(\omega)} < 1$.
By the triangle inequality,
\begin{equation}
\label{eq:triangle}
\begin{aligned}
    \lVert(W_\alpha^\text{loc} &+ W_\alpha^\text{nl}) R_H(\omega)\rVert \\
    \leq & \norm{ W_\alpha^\text{loc} R_H(\omega) } + \norm{ W_\alpha^\text{nl} } \, \norm{ R_H(\omega) }.
\end{aligned}
\end{equation}
For the first term, since $\omega$ is in the resolvent set, $R_H(\omega)$ maps into the domain of $H$ and the relative bound~\eqref{eq:rel-bound} applies. Using $H R_H(\omega) = -1 + \omega R_H(\omega)$, we find
\begin{equation}
\begin{aligned}
    \norm{ W_\alpha^\text{loc} R_H(\omega) }  
    \leq \alpha \Big[2a_0\left(1 + |\omega| \, \norm{ R_H(\omega) }\right) \\
     + (2b_0 + \alpha) \norm{ R_H(\omega) }\Big],
\end{aligned}
\end{equation}
which vanishes as $\alpha \to 0$.
The second term in~\eqref{eq:triangle} likewise vanishes since $\norm{ W_\alpha^\text{nl} } \to 0$.
We conclude that for sufficiently small $\alpha$, the resolvent $(\omega - \wt{H}_\alpha)^{-1}$ is bounded. 
This completes the proof.

\bibliography{bibliography}
\end{document}